# Study on the Thickness Change of Nickel-Plated Layer in Fabrication of the Silver Hollow Nickel Waveguides by the Outer-Coating Method of the Liquid Phase Process.

Sok-Bong Ro, Song-Jin Im and Uo-Hyon Kim

*Department of Physics,* **Kim Il Sung** *University, Pyongyang, DPR of Korea*

*Abstract －　A metallic hollow waveguide is promising fiber for the delivery of $CO_2$ laser radiation. Thickness of the nickel-plated layer for supporting of the waveguide in fabrication of a dielectric–coated silver hollow nickel waveguide is very important factor. In this paper, the change characteristic in the thickness of the nickel-plated layer along the length of the silver-coated glass mandrel during fabricating the silver hollow nickel waveguide by the outer-coating method of the liquid phase process has been studied both experimentally and analytically. Waveguides with uniform thickness of the nickel-plated layer along the length of the silver-coated glass mandrel have been fabricated.*

*Index terms －hollow waveguides, outer-coating method, thickness of plating layer*

## Ⅰ. INTRODUCTION

Hollow waveguide is one of the most attractive means of delivering infrared and high power or high energy laser. Two methods for fabrication of a dielectric-coated metallic hollow waveguides have already been developed [1-8]. One is the outer-coating method [1-3] and the other is the inner-coating method [4-8]. Moreover, techniques to coat the silver and COP or FCP layers inside stainless steel hollow structure for medical application have been developed[11,12]. Attenuation measurements of various kinds of the pulsed laser radiations through hollow silica waveguides were obtained and output beam quality was also studied [13].

There are two in outer-coating method. One is the method using aluminium tube as mandrel[1].The method requires an RF-sputtering apparatus which is used to deposit dielectric materials and a silver layer on which metal tube is formed, and this makes the fabrication process complicated. The other is to use a glass capillary tube as mandrel[3]. The advantage of the method is that the same RF-sputtering technique as used in Ref [1] is not used at all. Also, because the glass capillary tube with very smooth surface is used as mandrel, the inner surface irregularity of the waveguide is reduced. The fabrication steps are as follows: a polyimide layer is coated on the outer surface of the glass capillary tube, a silver layer is deposited by using a silver mirror reaction method and a nickel layer is formed by using electroplating method. Finally, the mandrel is etched away by 10% HCl solution. The diameter and the length of the waveguide is 800 $\mu m$ and 20 $cm$, respectively. However, in the fabrication of the waveguide by this technique, the thickness change of the electroplating layer along the length of the silver-coated mandrel in the plating process for the fabrication of the silver hollow nickel waveguides has not been investigated yet. The thickness of the plating layer is very important for supporting of the waveguide. For supporting of the waveguide, it has been known that the thickness of the nickel-plated layer should be about 70-200 $\mu m$ [1].

In this paper, we have investigated only the change characteristic in thickness of the nickel-plated layer along the length of the silver-coated mandrel in the plating process for the fabrication of the silver hollow nickel waveguides by the outer-coating method of the liquid phase process. We have fabricated the hollow waveguides with uniform thickness of the nickel- plated layer along the length of the silver-coated mandrel.

## II. EXPERIMENTAL RESULTS AND ANALYSIS

We have fabricated silver hollow waveguides for the transmission of $CO_2$ laser by using the outer-coating method used in Ref[3]. The method for fabrication is following: A silver layer is coated on the outer surface of a glass capillary tube by using the silver mirror reaction and the nickel layer is coated on the silver layer by the method used in Ref[1]. Finally, the glass capillary tube is dissolved by fluoric acid. The outer diameters and lengths of the glass capillary tube used as a mandrel are $0.9\,mm$ and $60\,cm$, respectively. The anode for the plating is a cylinder nickel tube and a silver-coated glass capillary tube used as the cathode is arranged in center of the anode. The cathode terminal is connected only to one of both ends of the silver-coated mandrel for the nickel plating. In the experiment, it has been observed that the thickness of the nickel-plated layer decreases rapidly with the increase of the length of the mandrel from the cathode terminal to the end of mandrel during the plating. Fig.1 shows the change characteristic in the thickness of the nickel-plated layer versus the length of the waveguides fabricated. The thickness of the nickel-plated layer has been determined: diameters of the waveguide after plating are measured at intervals of $10\,cm$ along the length of it and differences between them and outer diameter(0.9 $mm$) of the mandrel before plating are calculated.

The change in the thickness of the nickel-plated layer along the length of the silver-coated mandrel (cathode) indicates that the current density on the cathode differs from point to point on it. The change in the current density could be explained as a result of decrease in the cathode potential with the increase of the length of the cathode from the terminal. The decrease in the cathode potential could be resulted from the increase in the electric resistance of the silver layer coated on outer-surface of the glass capillary tube. The increase in the electric resistance is due to very thin thickness of the silver layer [9,10], hence very small cross section of it.

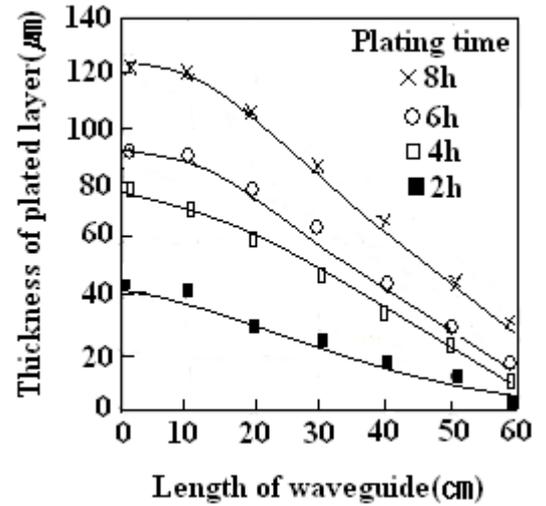

Fig.1. Relationship between the length of the waveguides and the thickness of the nickel-plated layer.

Fig.2 shows an experiment setup for the analysis of the change characteristic in the thickness of the nickel-plated layer. Firstly, we consider the case when the cathode terminal is connected only to A as shown in Fig.2 .

According to Faraday's electrolysis law, the mass of the plating material deposited on the cathode m is proportional to the intensity of current in the circuit $I$ through a plating solution and plating time $t$:

$$m = kIt \qquad (1)$$

where $k$ is a proportional factor. The cathode for a detailed consideration is divided into small segments.

The electric resistance $d\xi$ of any chosen segment $dx$ of a cathode should be given by

$$d\xi = \bar{\rho}\frac{dx}{\pi(r^2 - r_0^2)}, \qquad (2)$$

where $\bar{\rho}$ is resistivity of the silver, $r_0$ is a outer radius of a silver-coated glass capillary tube before plating and $r$ is a radius of any segment $dx$ after plating.

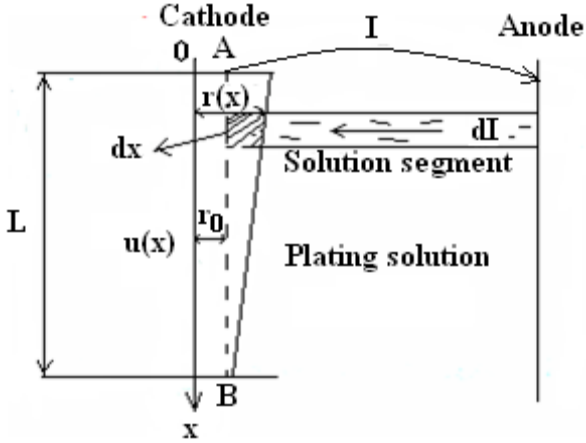

Fig.2. Experiment setup for the analysis of the thickness change of the nickel-plated layer.

Thus the electric resistance $\xi(x)$ for any interval $(0, x)$ is given as follow:

$$\xi(x) = \int_0^x d\xi = \frac{\overline{\rho}}{\pi} \int_0^x \frac{dx}{(r^2 - r_0^2)} \quad (3)$$

If the electric potential at the cathode terminal A is $u_0$ and the total current in the circuit is $I$, the potential at any position on the cathode is given by

$$u(x) = u_0 - I\xi(x). \quad (4)$$

Substituting (3) into (4), we obtain

$$u(x) = u_0 - I\frac{\overline{\rho}}{\pi}\int_0^x \frac{dx}{(r^2 - r_0^2)}. \quad (5)$$

Meanwhile, the plating solution could be regarded as a electric conductor. Therefore, the electric resistance of the plating solution would be given by

$$dR = \frac{\alpha}{dx}, \quad (6)$$

where $\alpha$ is a constant related to the characteristic of the solution.

Thus, the intensity of the current $dI$ through any segment of the plating solution would be given by

$$dI = \frac{u(x)}{dR} = \frac{dx}{\alpha}[u_0 - I\frac{\overline{\rho}}{\pi}\int_0^x \frac{dx}{(r^2 - r_0^2)}]. \quad (7)$$

According to Faraday's electrolysis law, the mass of the material $dm$ deposited on any segment of the cathode through plating solution is given by

$$dm = kdIt, \quad (8)$$

Where $t$ is a plating time.

On the other hand, the mass of the material $dm$ in the Eq.(8) could be written as

$$dm = \rho dV, \quad (9)$$

where $\rho$ is the density of the material plated on the cathode and $dV$ is the volume of the material deposited on any segment of the cathode:

$$dV = \pi(r^2 - r_0^2)dx. \quad (10)$$

Thus, we obtain following equation from (9) and (10):

$$dm = \rho\pi(r^2 - r_0^2)dx. \quad (11)$$

Substituting Eq. (7) and (11) into (8), following equation is obtained:

$$\rho\pi(r^2 - r_0^2)dx = \\ = kt\frac{dx}{\alpha}[u_0 - I\frac{\overline{\rho}}{\pi}\int_0^x \frac{dx}{(r^2 - r_0^2)}]. \quad (12)$$

Differentiating Eq. (12) with respect to $x$, we obtain following equations:

$$2\rho\pi rr' = -\frac{kt I \overline{\rho}}{\alpha\pi}\frac{1}{r^2 - r_0^2} \quad (13)$$

or

$$-\frac{2\alpha\rho\pi^2}{I\overline{\rho}k}\frac{r(r^2 - r_0^2)dr}{t} = dx, \quad (14)$$

where

$$\frac{2\alpha\rho\pi^2}{I\overline{\rho}k} \equiv c.$$

Substituting c into (14), we obtain following equation:

$$-\frac{c}{t}(r^3 - rr_0^2)dr = dx. \quad (15)$$

Integrating Eq.(15), the following equation is obtained:

$$-\frac{c}{t}(\frac{r^4}{4} - \frac{r^2 r_0^2}{2}) + D = x. \qquad (16)$$

This can lead to the following equation

$$-\frac{c}{4t}(r^4 - 2r^2 r_0^2) = x - D, \qquad (17)$$

where $D$ is the integration constant. Multiplying both sides of Eq.(17) by $\frac{4t}{c}$ we get equations:

$$r^4 - 2r^2 r_0^2 - \frac{4t}{c}(D - x) = 0 \qquad (18)$$

or

$$r^2 - r_0^2 = \sqrt{r_0^4 + \frac{4t}{c}(D - x)}. \qquad (19)$$

Eq. (18) and Eq. (19) are important in the analysis of the thickness change of the nickel-plated layer. Integration constant $D$ should be determined before the above equation could be solved. At $x = 0$ in Eq.(18) we get $r = r_1$, which is the maximum radius of the waveguide measured on the cathode terminal A as shown in Fig.2 after plating.

Thus, we obtain following equation:

$$r_1^4 - 2r_0^2 r_1^2 - \frac{4t}{c}D = 0. \qquad (20)$$

At $x = L$ in Eq.(18) we get $r = r_L$, which is minimum radius of the waveguide measured on B as shown in Fig.2 after plating.

Thus, we obtain

$$r_L^4 - 2r_0^2 r_L^2 - \frac{4t}{c}(D - L) = 0. \qquad (21)$$

From Eq. (20), we obtain

$$\frac{4t}{c} = \frac{r_1^2(r_1^2 - 2r_0^2)}{D}. \qquad (22)$$

Substituting Eq. (22) into (21), $D$ is given by

$$D = \frac{r_1^2(r_1^2 - 2r_0^2)L}{(r_1^4 - 2r_1^2 r_0^2) - (r_L^4 - 2r_L^2 r_0^2)}. \qquad (23)$$

Substituting Eq. (23) into (22) we obtain

$$\frac{4t}{c} = \frac{(r_1^4 - 2r_1^2 r_0^2) - (r_L^4 - 2r_L^2 r_0^2)}{L}. \qquad (24)$$

Substituting Eq. (23) and (24) into (19) we obtain

$$r^2 - r_0^2 = [r_0^4 + (r_1^4 - 2r_1^2 r_0^2)(1 - \frac{x}{L}) + (r_L^4 - 2r_L^2 r_0^2)\frac{x}{L}]^{1/2}. \qquad (25)$$

Thus, the radius $r$ of the waveguide after plating is obtained:

$$r = \{ r_0^2 + [ r_0^4 + (r_1^4 - 2r_1^2 r_0^2)(1 - \frac{x}{L}) + (r_L^4 - 2r_L^2 r_0^2)\frac{x}{L} ]^{1/2} \}^{1/2} \qquad (26)$$

Moreover the thickness of the nickel-plated layer $d$ is equal to the difference between the radius of the waveguide after plating and one of the mandrel before plating:

$$d = r - r_0 = \{ r_0^2 + [ r_0^4 + (r_1^4 - 2r_1^2 r_0^2)(1 - \frac{x}{L}) + (r_L^4 - 2r_L^2 r_0^2)\frac{x}{L} ]^{1/2} \}^{1/2} - r_0 \qquad (27)$$

The above equation represents the change characteristic in the thickness of the nickel-plated layer along the length of the waveguide from cathode terminal.

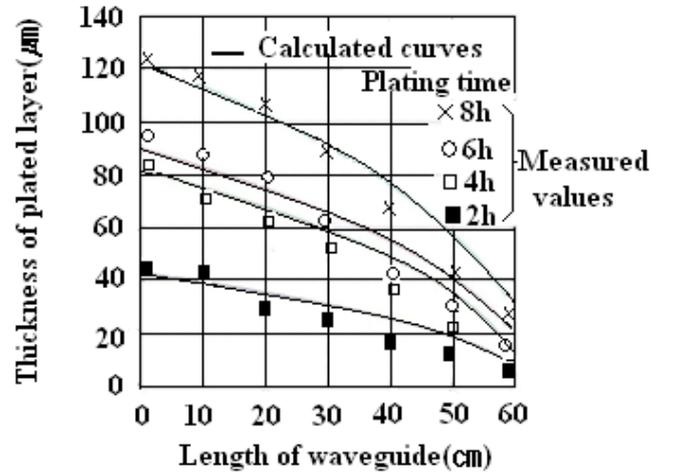

Fig.3 The thickness change of nickel-plated layer
(Calculated curves and measured values)

The Fig.3 shows the calculated curves from Eq. (27) and the measured thickness values of the nickel-plated layer along the length of the waveguide. In the Fig.3 there is good agreement between calculated curves from Eq.(27) and measured values from experiment. The same result is obtained when the cathode terminal is connected only to B as shown in Fig. 2,

From the above experimental and analytical results, it is concluded that the several cathode terminals should be arranged uniformly along the cathode for uniform thickness of the nickel-plated layer and decreased plating time. The thickness changes of the nickel-plated layer along the length of the waveguide and plating time are shown in Fig.4, where cathodes are connected to both of A and B. It is clear that the thickness of the nickel-plated layer is more uniform than that of case shown in the Fig.1.The plating time is also decreased.

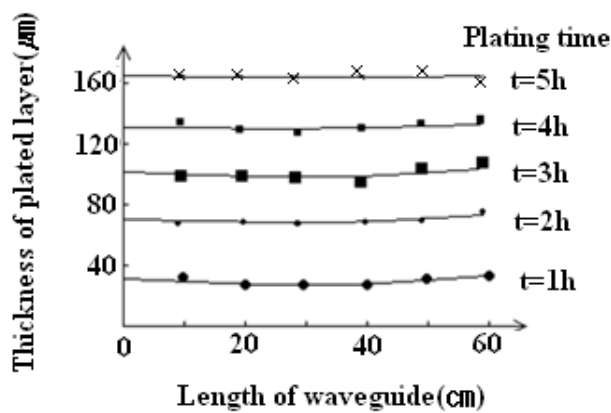

Fig.4 Thickness change of the plated layer (Here, cathode terminals are connected to both of A and B )

Ⅲ. CONCLUSIONS

In fabricating silver hollow nickel waveguide by the outer-coating method of the liquid phase process, the change characteristic in thickness of the nickel-plated layer has been studied both experimentally and analytically and the waveguides with uniform thickness of nickel-plated layer have been fabricated. The thickness of the nickel-plated layer needed for supporting the waveguide decreases rapidly as the length of the cathode from the cathode terminal increases. It showed a good agreement between experimental values and analytic results. The cathode terminals should be arranged uniformly for fabricating waveguides which has uniform nickel plated layer along the length of the mandrel.